\documentclass[preprint,12pt,nofootinbib]{revtex4-1}

\usepackage[brazil, english]{babel}
\usepackage{graphicx}
\usepackage{epsfig}
\usepackage{amssymb}
\usepackage{amsmath}
\usepackage{ulem}
\usepackage[T1]{fontenc} 
\usepackage[utf8]{inputenc}
\usepackage{color}
\usepackage{soul} 
\usepackage[titletoc]{appendix}

\begin{document}

\title{Holographic Description of Chiral Symmetry Breaking in a Magnetic Field in 2+1 Dimensions with an Improved Dilaton}
 
\author{Diego M. Rodrigues$^{1} $}
\email[Eletronic address: ]{diegomr@if.ufrj.br}
\author{Danning Li$^{2} $}
\email[Eletronic address: ]{lidanning@jnu.edu.cn}
\author{Eduardo Folco Capossoli$^{1,3}$}
\email[Eletronic address: ]{educapossoli@if.ufrj.br}
\author{Henrique Boschi-Filho$^{1}$}
\email[Eletronic address: ]{boschi@if.ufrj.br}  
\affiliation{$^1$Instituto de F\'\i sica, Universidade Federal do Rio de Janeiro, 21.941-972 - Rio de Janeiro-RJ - Brazil \\
$^2$Department of Physics and Siyuan Laboratory, Jinan University, Guangzhou 510632, China \\$^3$Departamento de F\'\i sica and Mestrado Profissional em Pr\'aticas da Educa\c c\~ao B\'asica (MPPEB), Col\'egio Pedro II, 20.921-903 - Rio de Janeiro-RJ - Brazil}

\begin{abstract}
We consider a holographic description of the chiral symmetry breaking in an external magnetic field in $ (2+1) $-dimensional gauge theories from the softwall model using an  improved dilaton field profile given by $\Phi(z) = - kz^2 + (k+k_1)z^2\tanh (k_{2}z^2)$. We find inverse magnetic catalysis for $B<B_c$ and magnetic catalysis for $B>B_c$, where $B_c$ is the pseudocritical magnetic field. The transition between these two regimes is a crossover and occurs at $B=B_c$, which depends on the fermion mass and temperature. We also find spontaneous chiral symmetry breaking (the chiral condensate $\sigma \not=0$) at $T=0$ in the chiral limit ($m_q\to 0$) and chiral symmetry restoration for finite temperatures. We observe that changing the $k$ parameter of the dilaton profile only affects the overall scales of the system such as $B_c$ and $\sigma$. For instance, by increasing $k$ one sees an increase of $B_c$ and $\sigma$. This suggests that increasing the parameters $k_1$ and $k_2$ will decrease the values of $B_c$ and $\sigma$. 
\end{abstract}


\maketitle

\newcommand{\limit}[3]
{\ensuremath{\lim_{#1 \rightarrow #2} #3}}

\section{Introduction}   

During recent years a lot of effort has been done in order to understand the interplay between a magnetic field and chiral phase transition. It has been long thought that a magnetic catalysis (MC) should occur in 2+1  dimensions \cite{Gusynin:1994re,Miransky:2002rp,Shovkovy:2012zn,Miransky:2015ava}, where the magnetic field boost the chiral condensate/transition temperature. However, undeniable lattice evidences for inverse magnetic catalysis (IMC) behavior (a decrease of the chiral condensate when the external magnetic field increases)  in $ 3+1 $ dimensions was presented in \cite{Bali:2011qj,Endrodi:2015oba} for $eB$ up to $\sim 3\, $GeV$^2$. 

With the advent of the holographic duality, originally called AdS/CFT correspondence \cite{Maldacena:1997re}, several attempts have been made in order to obtain such IMC transition behavior from holographic descriptions of QCD, also known as AdS/QCD models \cite{Preis:2010cq,Mamo:2015dea,Dudal:2015wfn,Li:2016gfn,Evans:2016jzo}.

The main goal of this work is to analyze the chiral symmetry breaking in the presence of an external magnetic field in 2+1 dimensions using the holographic softwall model with an improved dilaton profile \begin{equation} 
\Phi(z) = - kz^2 + (k+k_1)z^2\tanh (k_{2}z^2). \nonumber 
\end{equation} 
Such improvement means an interpolation between the IR and UV regimes of the dual field theory, which represents a UV completion with respect to the standard dilaton $\Phi(z) = - kz^2$, as used in \cite{Rodrigues:2018pep} to study the chiral phase transition and spontaneous chiral symmetry breaking in the presence of an external magnetic field. 
Many works dealt with modified dilaton fields to implement UV completion in different contexts as discussed for instance in  \cite{Shock:2006gt, White:2007tu, Pirner:2009gr, Mia:2010tc, He:2010ye, Li:2011hp,Li:2013oda}.

This modification of the dilaton profile was crucial to correctly reproduce the spontaneous chiral symmetry breaking, chiral phase transition and IMC in holographic QCD in 3+1 dimensions \cite{Chelabi:2015cwn,Chelabi:2015gpc,Li:2016gfn}. Recently, another improved dilaton profile has also been used to describe the dissociation of heavy mesons in a plasma with magnetic fields \cite{Braga:2018zlu}.

Here, we describe holographically with an improved dilaton the behaviour of the chiral condensate under the presence of an external magnetic field at $T=0$ and also at finite temperature ($T>0$). We find IMC for $B<B_c$ and MC for $B>B_c$, where $B_c$ is the pseudo critical magnetic field associated with the crossover transition. We also find spontaneous chiral symmetry breaking at $T=0$ in the chiral limit ($m_q\to0$) and chiral symmetry restoration at finite temperature. Since the present model is more robust than the standard dilaton one can infer that the improved dilaton profile gives support to the results presented in \cite{Rodrigues:2018pep}. 

This work is organized as follows: in section \ref{sec2} we describe the holographic set up for chiral symmetry breaking in the presence of an external magnetic field in 2+1 dimensions.  
In section III, we present our numerical results concerning the behaviour of the chiral condensate versus magnetic field and temperature. Finally, in section IV, we present our last comments and conclusions. 


\section{Holographic softwall model with improved dilaton: chiral symmetry breaking}
\label{sec2}

\begin{figure}[!h]
	\centering
	\includegraphics[scale = 0.24]{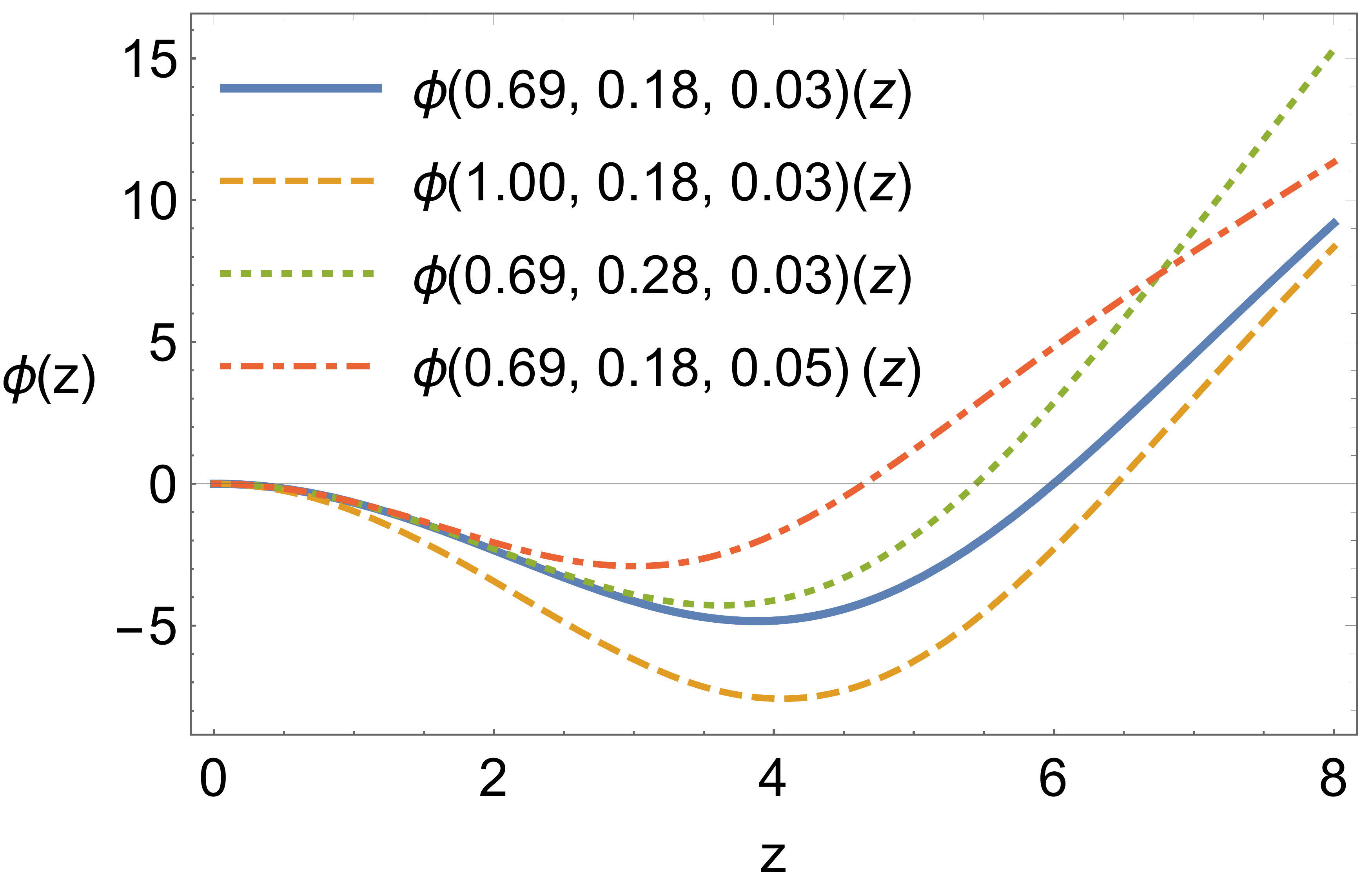}
	\caption{The dilaton profile eq. \eqref{dilaton profile} for some choices of the parameters  $k$, $k_1$, and $k_2$ for $\Phi(k, k_1, k_2)(z)$. The solid line represents the choice of parameters used in this work.}
\label{fig_dilaton}
\end{figure}

\begin{figure}[!h]
	\centering
	\includegraphics[scale = 0.65]{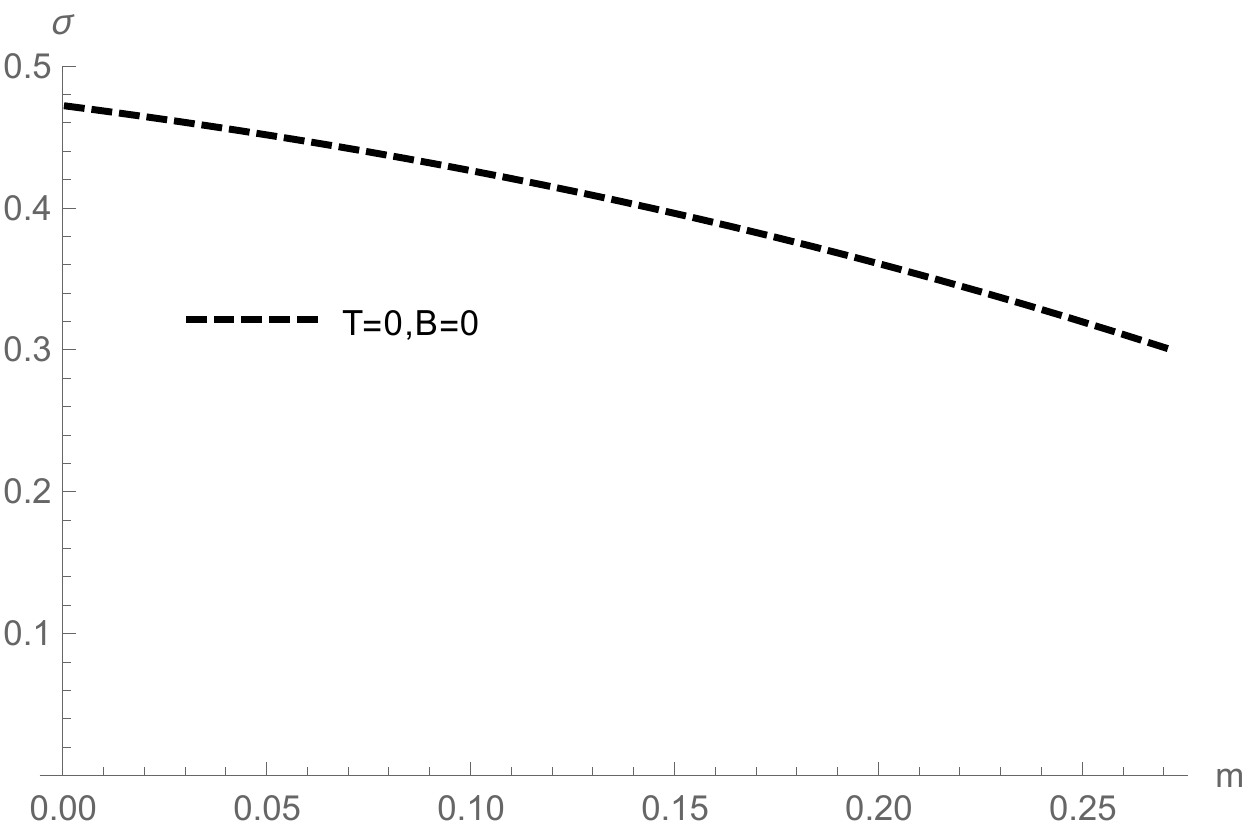}
	\caption{This figure represents the chiral condensate $\sigma$, in units of ${\sigma_s}$, against the fermion mass $m_q$, in units of ${\sqrt{\sigma_s}}$. In this figure one can clearly see spontaneous chiral symmetry breaking in the chiral limit $m_q\rightarrow0$.}

\label{fig_massa}
\end{figure}

\begin{figure}[!h]
	\centering
	\includegraphics[scale = 0.55]{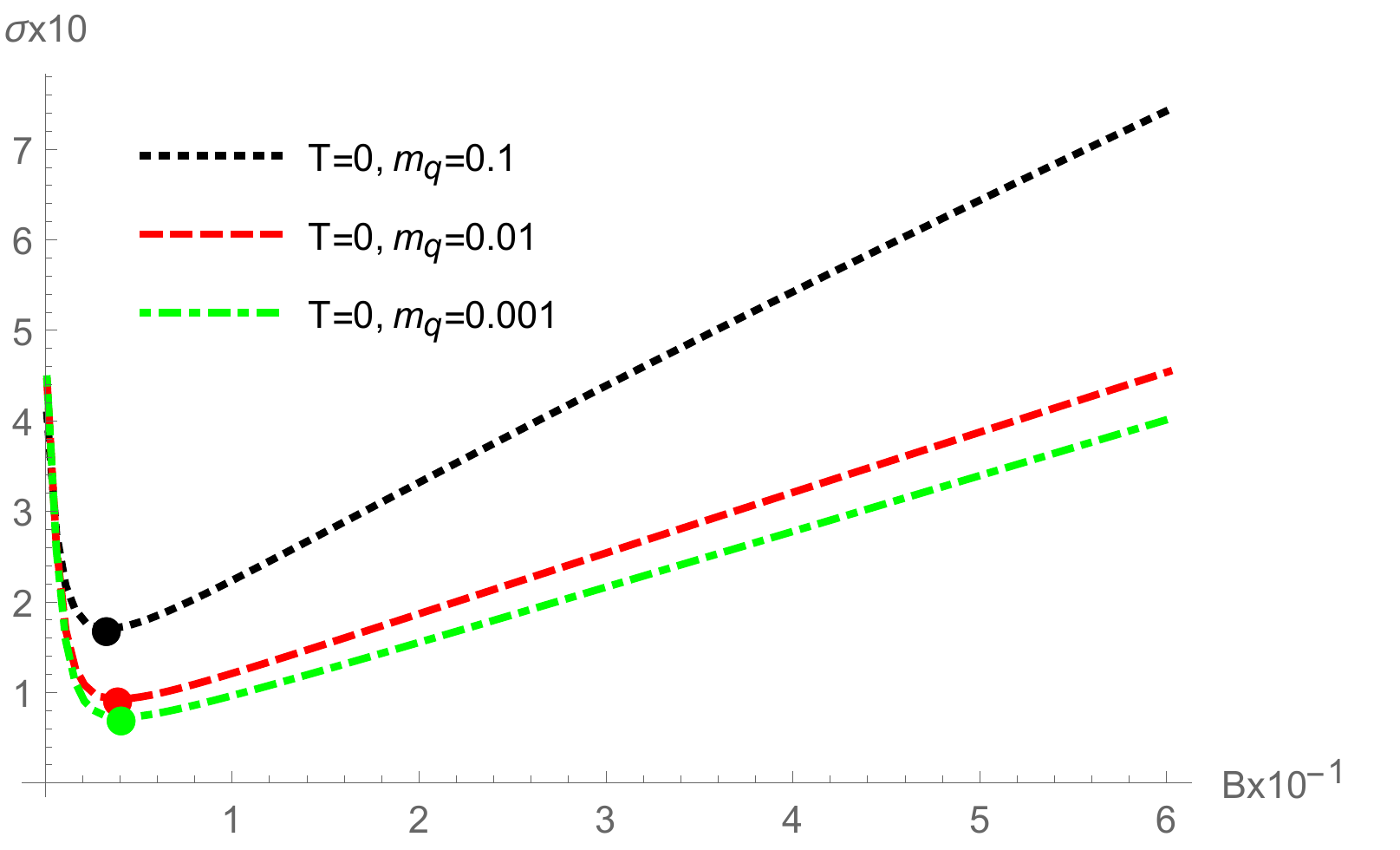}
	\caption{This figure represents the chiral condensate $\sigma$ against the magnetic field $B$, both in units of ${\sigma_s}$ at zero temperature, and three fermion masses. This figure shows the IMC phase for $B<B_c$ and the MC phase for $B>B_c$, where $B_c$  is the pseudocritical magnetic field pointed out by the colored disks. The smooth transition between the IMC and MC phases is a crossover. The values found for $B_c/{\sigma_s}$ are: 3.33, 3.93, and 4.09  for ${m_q}/{\sqrt{\sigma_s}}$ = 0.1, 0.01, and 0.001, respectively.   
}
	\label{fig_1}
\end{figure}

\begin{figure}[!h]
	\centering
	\includegraphics[scale = 0.55]{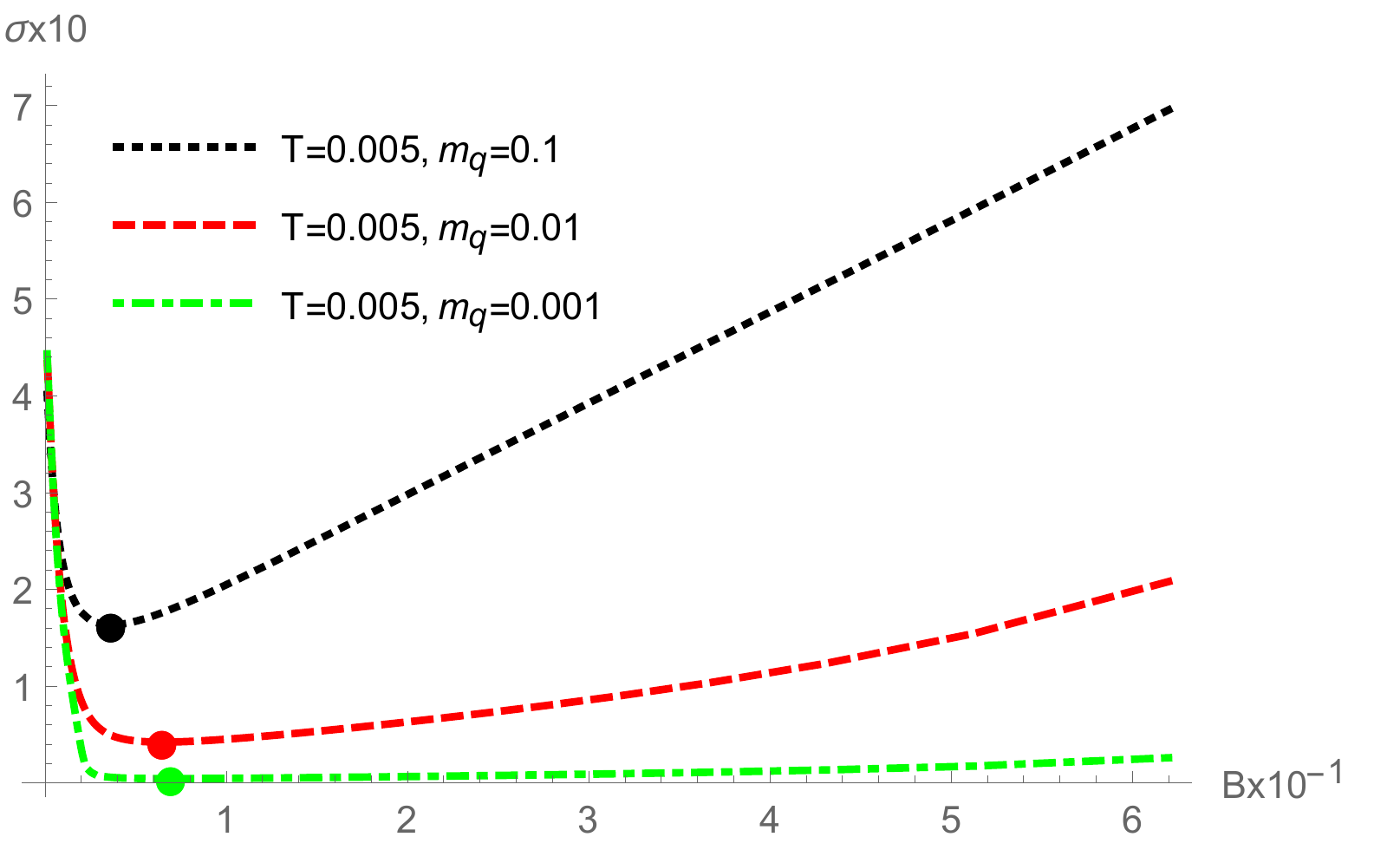}
	\caption{This figure represents the chiral condensate $\sigma$ against the magnetic field $B$, both in units of ${\sigma_s}$ at $T =0.005$, in units of ${\sqrt{\sigma_s}}$, and three fermion masses. This figure shows the IMC phase for $B<B_c$ and the MC phase for $B>B_c$, where $B_c$  is the pseudocritical magnetic field pointed out by the colored disks. The smooth transition between the IMC and MC phases is a crossover. The values found for $B_c/{\sigma_s}$ are: 3.65, 6.47, and 6.94  for ${m_q}/{\sqrt{\sigma_s}}$ = 0.1, 0.01, and 0.001, respectively. 
}.
	\label{fig_2}
\end{figure}

\begin{figure}[!h]
	\centering
	\includegraphics[scale = 0.55]{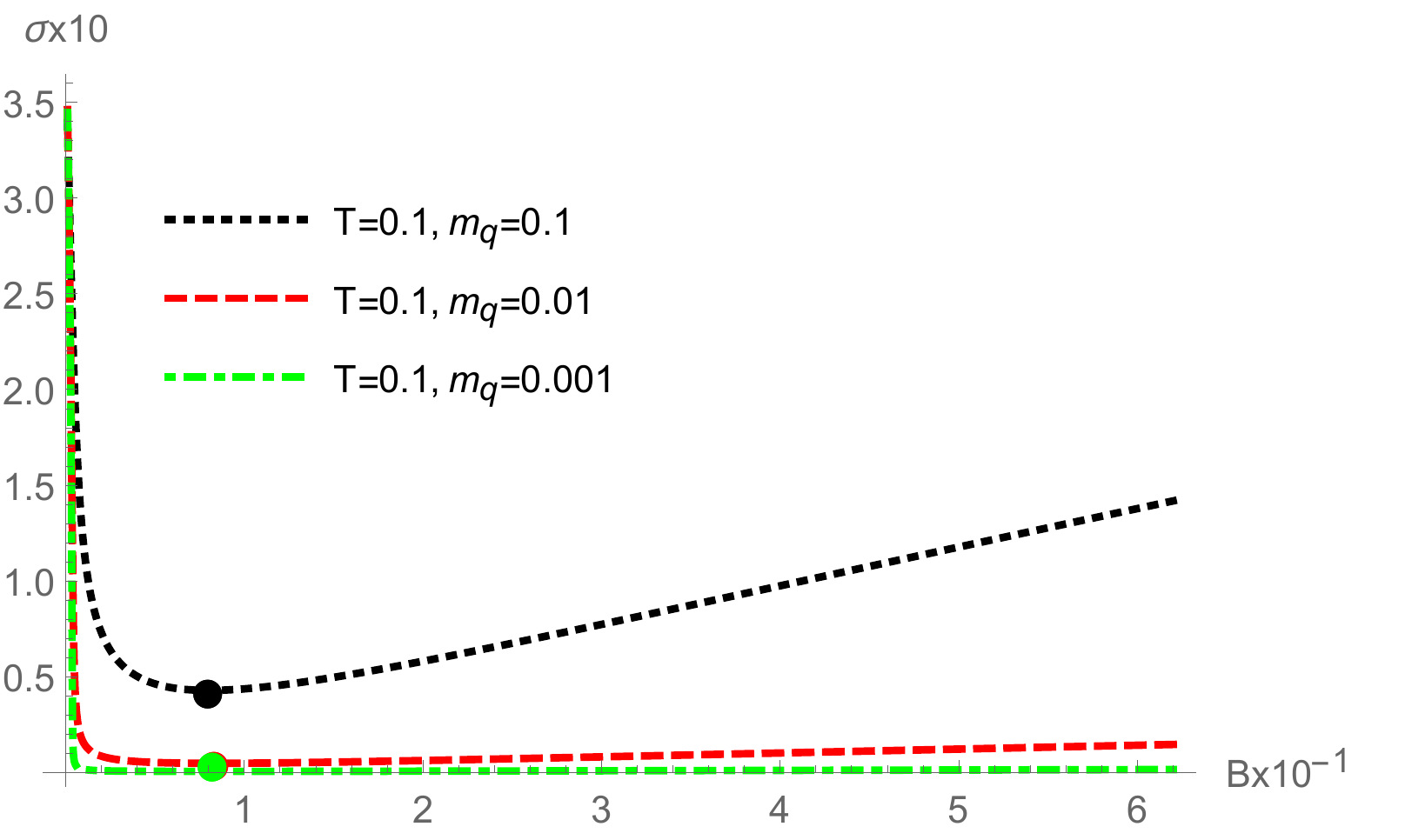}
	\caption{This figure represents the chiral condensate $\sigma$ against the magnetic field $B$, both in units of ${\sigma_s}$ at $T =0.1$, in units of ${\sqrt{\sigma_s}}$, and three fermion masses. This figure shows the IMC phase for $B<B_c$ and the MC phase for $B>B_c$, where $B_c$  is the pseudocritical magnetic field pointed out by the colored disks. The smooth transition between the IMC and MC phases is a crossover. The values found for $B_c/{\sigma_s}$ are: 8.00, 8.34, and 8.26  for ${m_q}/{\sqrt{\sigma_s}}$ = 0.1, 0.01, and 0.001, respectively. 
}
	\label{fig_3}
\end{figure}

\begin{figure}[!h]
	\centering
	\includegraphics[scale = 0.55]{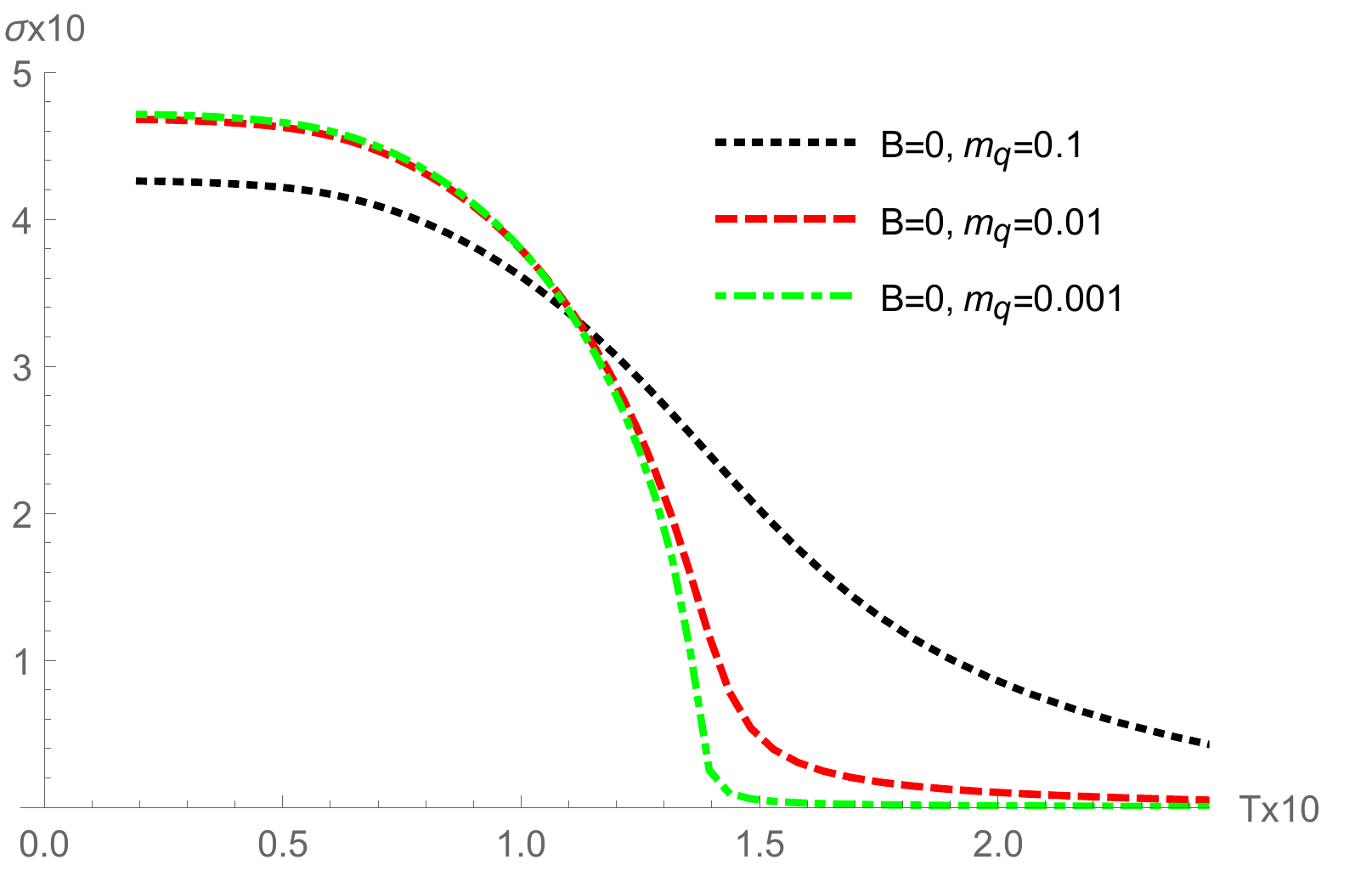}
	\caption{This figure shows the chiral condensate $\sigma$, in units of ${\sigma_s}$, versus the temperature $T$, in units of $\sqrt{\sigma_s}$. Here one can see the chiral symmetry restoration ($\sigma(T)=0$) for different values of the fermion mass without external magnetic field. One can also see the unexpected result that $\sigma(B=0)\neq0$ for low temperatures. 
}
	\label{fig_4}
\end{figure}

\begin{figure}[!h]
	\centering
	\includegraphics[scale = 0.55]{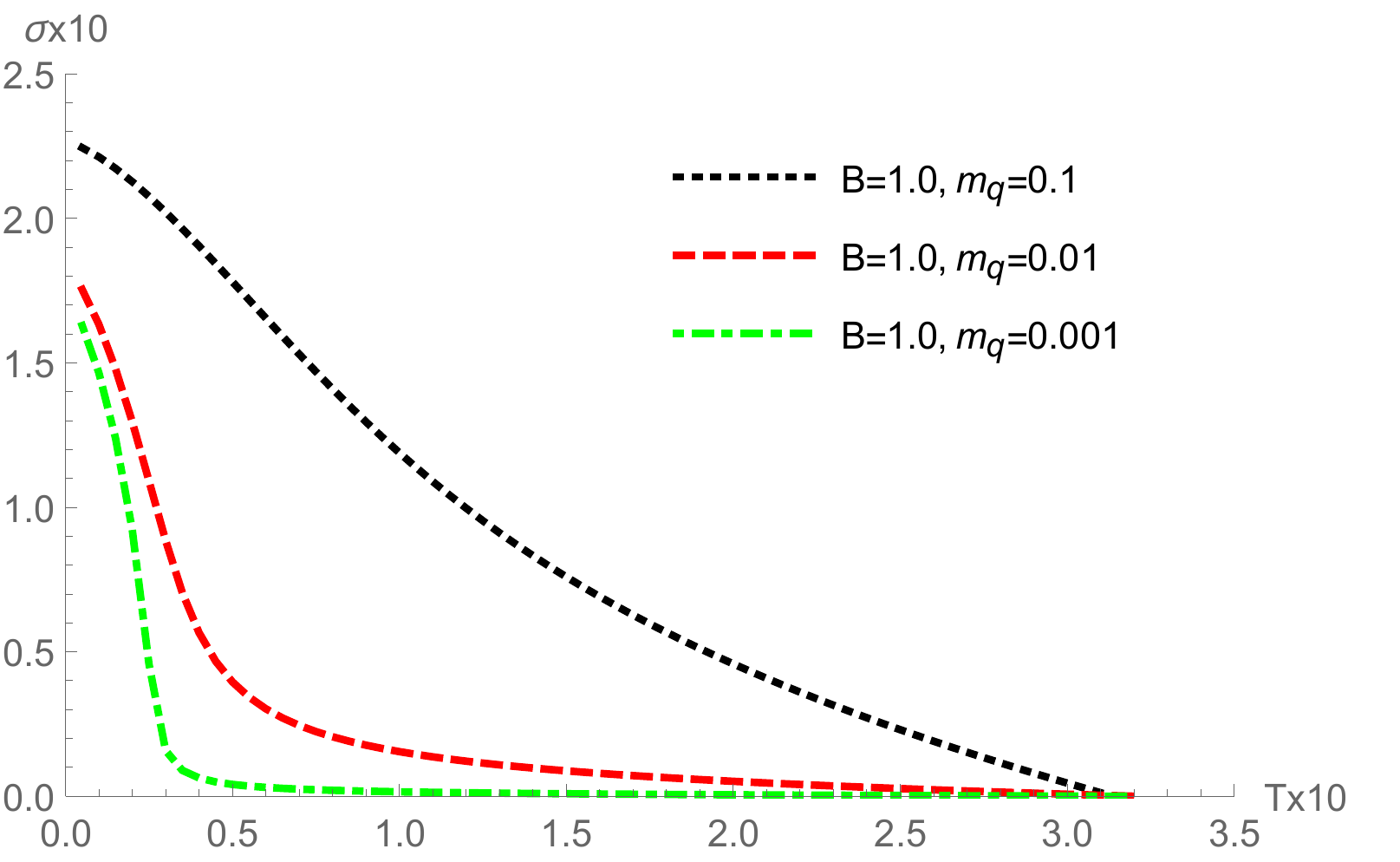}
	\caption{This figure shows the chiral condensate $\sigma$, in units of ${\sigma_s}$, versus the temperature $T$, in units of $\sqrt{\sigma_s}$. Here one can see the chiral symmetry restoration ($\sigma(T)=0$) for different values of the fermion mass in the presence of an external magnetic field ${B}/{\sigma_s} = 1.0$. Here one can see the chiral symmetry restoration ($\sigma(T)=0$) for different values of the fermion mass. 
}
	\label{fig_5}
\end{figure}

\begin{figure}[!h]
	\centering
	\includegraphics[scale = 0.55]{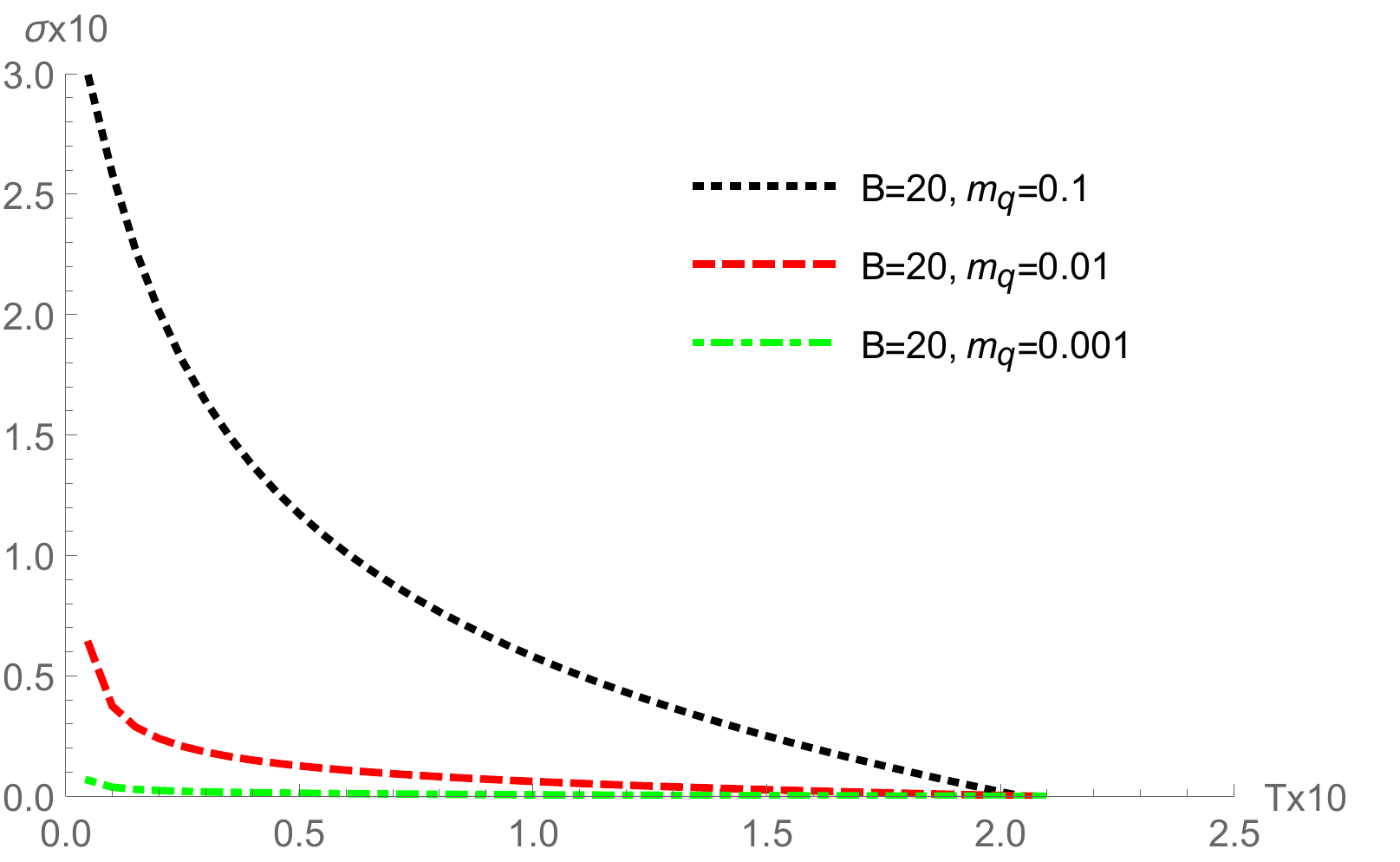}
	\caption{This figure shows the chiral condensate $\sigma$, in units of ${\sigma_s}$, versus the temperature $T$, in units of $\sqrt{\sigma_s}$. Here one can see the chiral symmetry restoration ($\sigma(T)=0$) for different values of the fermion mass in the presence of an external magnetic field ${B}/{\sigma_s} = 20$. Here one can see the chiral symmetry restoration ($\sigma(T)=0$) for different values of the fermion mass. 
}
	\label{fig_6}
\end{figure}

\begin{figure}[!h]
	\centering
	\includegraphics[scale = 0.55]{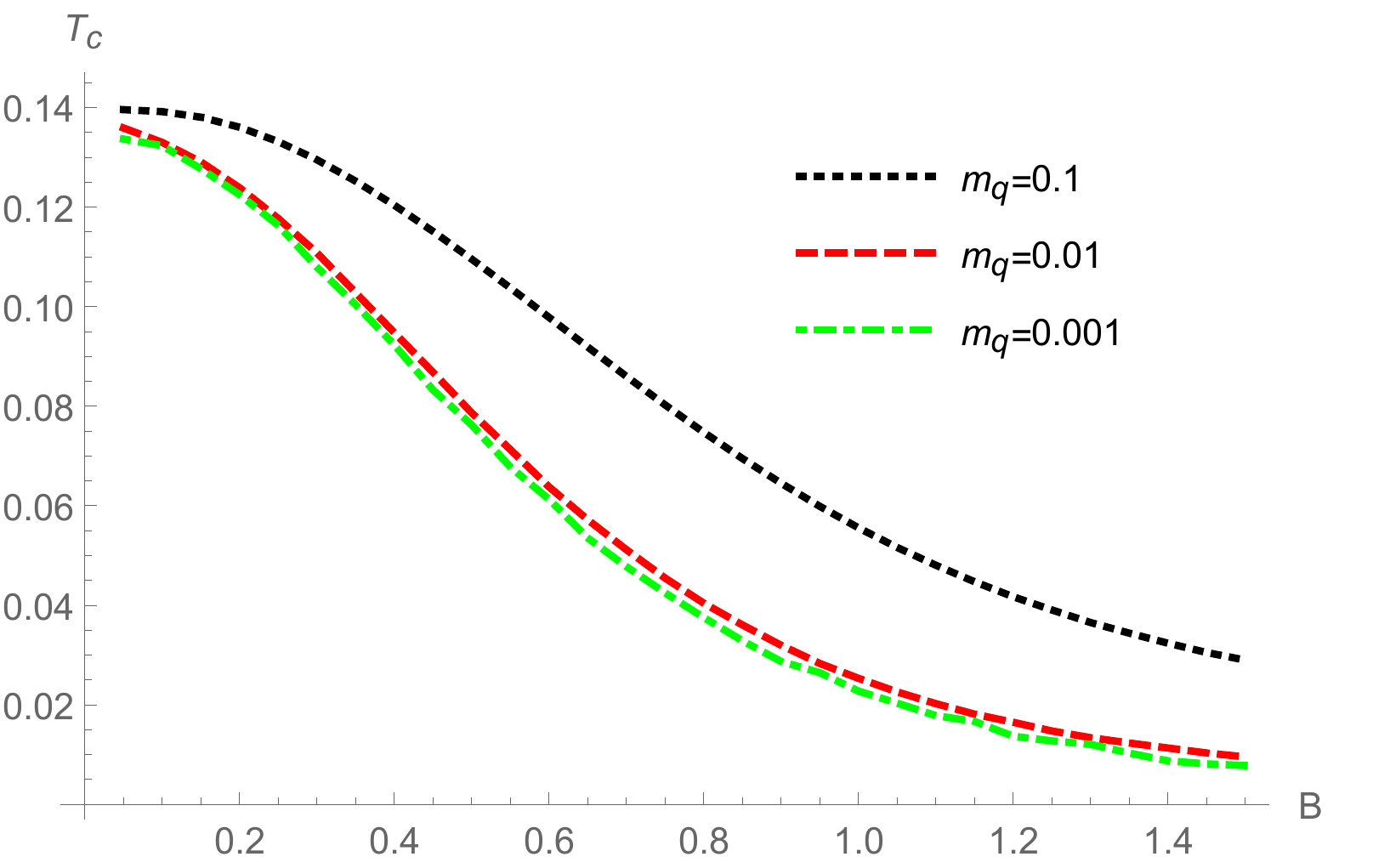}
	\caption{This figure shows the critical temperature, $T_c$, in units of $\sqrt{\sigma_s}$, against the magnetic field, $B$, in units of $ \sigma_s $, in the IMC phase,  for different values of the fermion mass.   
}
	\label{fig_7}
\end{figure}

\begin{figure}[!h]
	\centering
	\includegraphics[scale = 0.55]{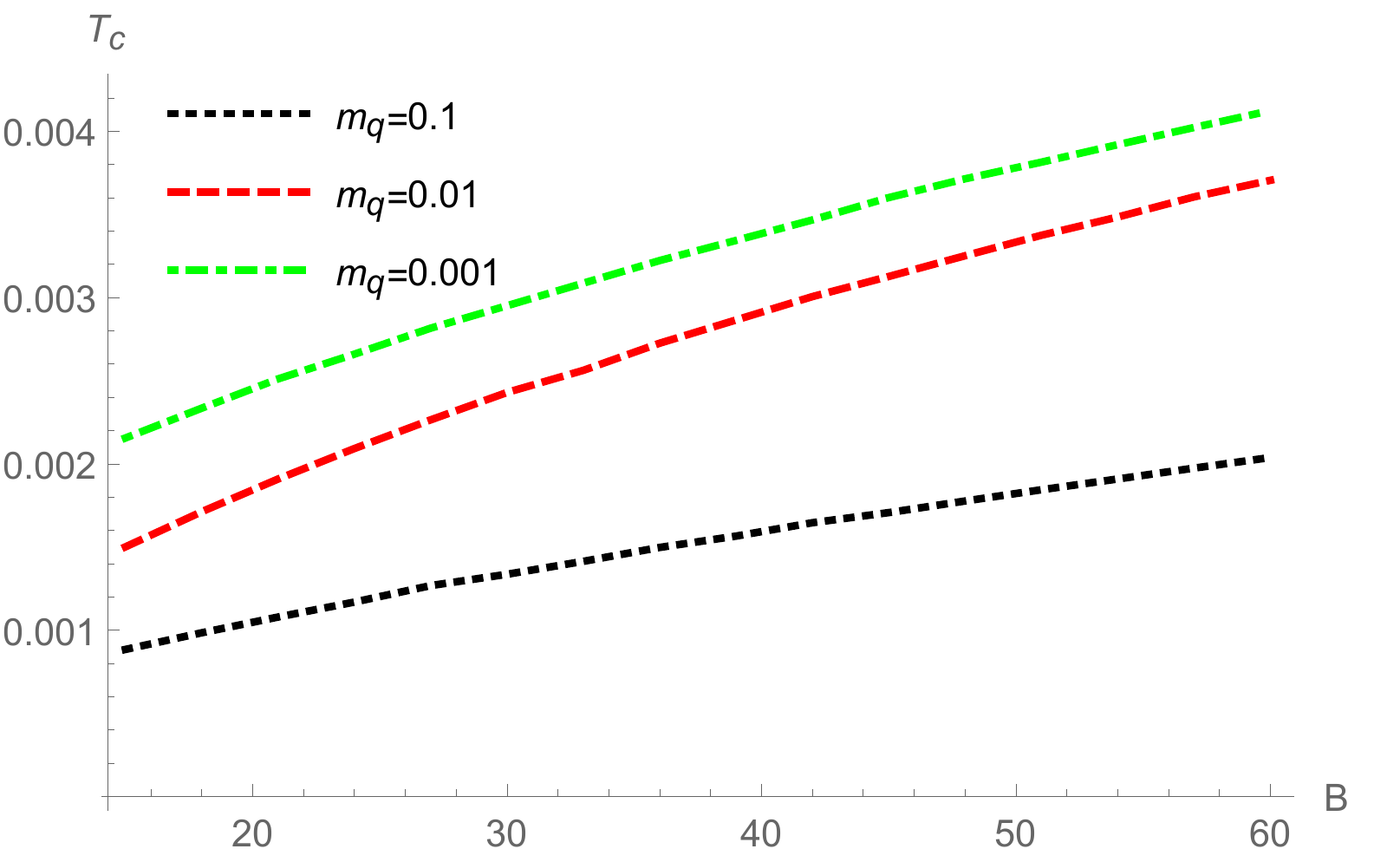}
	\caption{This figure shows the critical temperature, $T_c$, in units of $\sqrt{\sigma_s}$, against the magnetic field, $B$, in units of $ \sigma_s $, in the MC phase, for different values of the fermion mass.   
}
	\label{fig_8}
\end{figure}

Let us begin this section with a quick review of the holographic softwall model (SW). This model   successfully breaks the conformal invariance coming from the AdS/CFT correspondence. Such conformal invariance is broken by using an  exponential factor representing the dilaton field in the action producing a soft IR cutoff in the dual gauge theory.

The original SW model was proposed in \cite{Karch:2006pv} to study mesonic spectra and produce a linear Regge trajectory. This model was extended to the case of glueballs in \cite{Colangelo:2007pt}. 
In ref. \cite{Capossoli:2015ywa}, based on a dynamical and analytic modified holographic softwall model, it was shown that the original SW seems to not work properly for glueballs when compared to lattice data and other approaches in the literature. For other details see for instance \cite{FolcoCapossoli:2016ejd} 
and references therein. 

After this very brief review, let us start our calculation within the $AdS_4/CFT_3$ version of correspondence. So, the Einstein-Maxwell theory on $AdS_4$ \cite{Herzog:2007ij} is represented by the following action (for more details see \cite{Rodrigues:2017cha,Rodrigues:2017iqi}):
\begin{equation} \label{AdS4Action}
S = -\dfrac{1}{2\kappa^2_4}\int d^{4}x \sqrt{g}\left(\mathcal{R} -2\Lambda - L^{2}F_{MN}F^{MN}\right),
\end{equation}
\noindent where $\mathcal{R} = -12/L^2$ is the Ricci scalar, $\Lambda = -3/L^2$ is the cosmological constant and $F_{MN}$ is the Maxwell field. For our purposes, throughout the text we will use the AdS radius $L=1$.

Varying the action \eqref{AdS4Action} with respect to the fields, one gets: 
\begin{equation}\label{FieldEquations}
R_{MN} = 2\left( F_{M}^{P}F_{NP}-\dfrac{1}{4}g_{MN}F^2\right)  - 3g_{MN}, 
\end{equation}
plus the Bianchi identity: $\nabla_M F^{MN}=0$.

In order to solve \eqref{FieldEquations} let us consider the $AdS_4$-Schwarzschild  metric: 
\begin{equation}\label{AnsatzMetrica} 
ds^2 = z^{-2}\left( f(z)d\tau^2 + \dfrac{dz^2}{f(z)} + dx^2_1 + dx^2_2\right), 
\end{equation}
\noindent where $f(z)$ is horizon function. Since this is a charged black hole it has two horizons, the inner and the outer. Here, only the outer one, which satisfies $f'(z=z_H) < 0$, will be relevant for our analysis.
For the Maxwell field, the ansatz we consider is the following   
\begin{equation}\label{magfield}
F = B\, dx_{1}\wedge dx_{2},
\end{equation}
representing an uniform magnetic field in the $z$ direction, 
such that $F = dA$, where $A$ is the 1-form vector potential
\begin{equation}
A = \frac{B}{2}(x_1 dx_2 - x_2dx_1),
\end{equation}
which is nonzero ($A\neq0$) at the boundary ($z\to0$).

Substituting Eqs. \eqref{AnsatzMetrica} and \eqref{magfield} in \eqref{FieldEquations}, one finds that the horizon function, which satisfies $ f(z_H)=0 $, is given by \cite{Rodrigues:2017cha,Rodrigues:2017iqi}
\begin{eqnarray}
f(z) &=& 1 + B^{2}z^3(z-z_H) - \dfrac{z^3}{z^{3}_H}\,, \label{fBT}
\end{eqnarray} 
for the outer horizon. The corresponding temperature is given by the Hawking formula
$T = {|f'(z=z_H)|}/{4\pi}$. 
Using the solution \eqref{fBT} and the condition $f'(z=z_H) < 0$, we have: 
\begin{equation}
T(z_{H},B) = \dfrac{1}{4\pi}\left( \dfrac{3}{z_{H}} - B^2 z_{H}^3 \right),  \quad z_{H}^4< \dfrac 3 {B^2}. 
\end{equation}

The chiral symmetry breaking in the softwall model is described by the action \cite{Chelabi:2015cwn,Chelabi:2015gpc}: 
\begin{equation}\label{ChiralAction}
S = -\dfrac{1}{2\kappa^2_4}\int d^{4}x \sqrt{g}\,e^{-\Phi(z)}\mathrm{Tr}\left(D_{M}X^{\dagger}\,D^{M}X + V_{X} - F_{MN}^2 \right), 
\end{equation}
where $ X $ is a complex scalar field dual to the chiral condensate $ \sigma\equiv\left\langle \bar{\psi}\psi\right\rangle $ in $ 2+1 $ dimensions, $ D_{M} $ is the covariant derivative, $ F_{MN}$ is the field strength and $ V_{X} = -2X^2 + \lambda X^4 $ is the non-linear interaction necessary to realize the spontaneous symmetry-breaking mechanism \cite{Gherghetta:2009ac,Chelabi:2015cwn,Chelabi:2015gpc}. The  equations of motion coming from \eqref{ChiralAction} are given by
\begin{equation}\label{ChiralFieldEquations}
D_M\left[ \sqrt{g} \;  e^{-\Phi(z)} g^{MN} D_N {X}\right] - \sqrt{g} 
e^{-\Phi(z)}\partial_{X}V_X = 0,
\end{equation}
where the dilaton field takes the improved form 
\begin{equation} \label{dilaton profile}
\Phi(z) = - kz^2 + (k+k_1)z^2\tanh (k_{2}z^2),
\end{equation}
which at the UV regime ($z\to 0$) gives $ \Phi(z) = -kz^2 $, while in the IR limit ($z\to \infty$) we have $ \Phi(z) = k_1 z^2 $, with $ k $,  $ k_1 $ and $ k_2 $ being constants to be fixed later. 

Assuming that $\langle X(x^{\mu},z) \rangle \propto \chi(z) $ one can write \eqref{ChiralFieldEquations} as \cite{Dudal:2015wfn,Li:2016gfn}
\begin{equation}\label{ChiralFieldEquations2}
\chi''(z) + \left(-\frac{2}{z} - \Phi'(z)+ \frac{f'(z)}{f(z)}\right) \chi'(z) - \dfrac{1}{z^2 f(z)}\partial_{\chi}V(\chi) = 0,
\end{equation}
where prime denotes derivative with respect to $ z $. 

In this work we solve numerically Eq. \eqref{ChiralFieldEquations2} for zero and finite temperature using the quartic potential $ V(\chi) = -\chi^2+\lambda\chi^4 $, with $ \lambda=1$. The boundary conditions used are: (i) at the UV $\chi(z) = m_{q}z + \sigma z^2$, and (ii) the regularity of $ \chi(z) $ at the horizon, $ \chi(z_{H})<\infty $ \cite{Rodrigues:2018pep}. Since we are working in 2+1 dimensions all dimensionful  parameters like $k$, $\sigma$, $B$, $T$, and the fermion mass $m_{q}$ will be measured in units of the string tension ($\sqrt{\sigma_s}$). So, the temperature and mass will be measured in units of $\sqrt{\sigma_s}$, as for instance in refs. \cite{Teper:1998te, Meyer:2003wx, Athenodorou:2016ebg}. On the other side, the magnetic field and the chiral condensate will be measured in units of the string tension squared $(\sqrt{\sigma_{s}})^2$. 

\section{Numerical Results}

In this section we present our results for the chiral symmetry breaking using the holographic softwall model with an improved dilaton field given by \eqref{dilaton profile}. For our numerical analysis we choose $k = 0.69$, $k_1= 0.18  $, and $k_2= 0.031 $, all in units of $\sigma_s$. The values for these parameters were taken from refs. \cite{Chelabi:2015cwn, Chelabi:2015gpc} which deal with this problem in 3+1 dimensions. There the value of $k$ comes from the mass of the rho meson, and $k_1$ and $k_2$ were chosen to fit lattice data of the critical temperature and the chiral condensate. In 2+1 dimensions these data are not available. In Fig. \ref{fig_dilaton} we plot the dilaton profile for some values of the parameters $k$, $k_1$, and $k_2$. When we increase $k$ we see that the curve deepens, while for $k_1$ and $k_2$ the opposite happens together with the dislocation of the minimum to the region of small $z$.

In Figure \ref{fig_massa} we show the behavior of the chiral condensate $\sigma$, in units of $\sigma_s$, against the fermion mass $m_q$, in units of $\sqrt{\sigma_s}$. One can see a finite value of $\sigma$ in the chiral limit ($m_q\rightarrow0$) which characterizes a spontaneous chiral symmetry breaking. Note that the value of the condensate diminishes for increasing fermion mass. This is in contrast with the perturbative result \cite{Gusynin:1994re,Miransky:2002rp,Shovkovy:2012zn,Miransky:2015ava}. Note that our analysis is non-perturbative in nature. This behavior can also be clearly seen, for instance, in Fig. \ref{fig_4}, but it disappears when one increases the temperature or the value of the magnetic field.

In Figures \ref{fig_1}, \ref{fig_2}, and \ref{fig_3} we show the behavior of the chiral condensate $\sigma$ against the external magnetic field $B$, for three different quark masses $m_q$ and three different temperatures $T=0$, $T=0.005$, and $T=0.1$, in units of $\sqrt{\sigma_s}$. In these three pictures, one can see for weak magnetic fields ($B<B_c$) a behavior known as IMC, where $B_c$ is the pseudocritical field (indicated by colored disks). For strong fields ($B>B_c$) one finds MC. These pictures also show that the transition between these two regimes is a crossover. 

In Figures \ref{fig_4}, \ref{fig_5} and \ref{fig_6}  we show the behavior of the chiral condensate $\sigma$ against the temperature $T$, for three different quark masses $m_q$ and three different magnetic fields $B=0$, $B=1.0$, and $B=20$, in units of the string tension squared ${\sigma_s}$. In these three pictures, one can see that the chiral condensate decreases as the temperature increases. This behavior is consistent with chiral symmetry restoration. In particular, in Fig. \ref{fig_4}, for $B = 0$ we have a nonzero chiral condensate ($\sigma(B=0)\neq0$) for low temperatures. Such behavior does not appear neither in perturbative results in 2+1 dimensions  \cite{Gusynin:1994re, Miransky:2002rp, Shovkovy:2012zn, Miransky:2015ava} nor in lattice QCD in 3+1 dimensions \cite{Bali:2011qj,Endrodi:2015oba}.

In Figures \ref{fig_7} and \ref{fig_8} we show the behavior of the critical temperature $T_c$, in units of  $\sqrt{\sigma_s}$ against the magnetic field $B$, in units of  ${\sigma_s}$, for three different fermion masses $m_q$. The Figure \ref{fig_7} represents the behavior of the critical temperature $T_c$ for weak magnetic fields in the range $ 0.1 \leq B/{\sigma_s} \leq 1.5$. On the other hand, Figure \ref{fig_8} represents the behavior of the critical temperature $T_c$ for strong magnetic fields in the range $ 15 \leq B/{\sigma_s} \leq 60$.


\section{Conclusions}
Here in this work we studied the holographic description of chiral symmetry breaking in the presence of an external magnetic field in 2+1 dimensions using the softwall model with an improved dilaton profile, given by Eq. \eqref{dilaton profile}. This profile interpolates two different behaviors in UV and IR: for $z\to 0$ (UV) one has $ \Phi(z) = -kz^2 $, on the other hand, for $z\to \infty$ (IR) one has   $ \Phi(z) = k_1 z^2 $. Furthermore, this dilaton profile has been designed to conform with the confinement criteria in the IR which establishes that it should go as $k_1 z^2$ (positive) as $z\to \infty$ \cite{Gursoy:2007cb, Gursoy:2007er, Gursoy:2008za}.

From this study we obtained spontaneous chiral symmetry breaking in the chiral limit ($m_q\to0$) as can be seen in Figure \ref{fig_massa}. From the chiral condensate as function of the external magnetic field we have shown the IMC/MC transition, which is a crossover, separated by a pseudocritical magnetic field $B_c$, as can be seen in Figures  \ref{fig_1}, \ref{fig_2}, and \ref{fig_3}. 
 
From the Figures \ref{fig_4}, \ref{fig_5}, and \ref{fig_6} which represent the behaviour of the chiral condensate against temperature one can see the restoration of the chiral symmetry.

The critical temperature as a function of the external magnetic field,  also pointing out IMC and MC, was presented in Figures \ref{fig_7}  and \ref{fig_8}, respectively.

The results presented in this work give support to the ones found previously in \cite{Rodrigues:2018pep} with a negative quadratic dilaton in 2+1 dimensions.
This suggests the robustness of the softwall model with the improved dilaton to describe the chiral symmetry breaking and chiral phase transition since it works in 3+1 dimensions \cite{Li:2016gfn, Chelabi:2015cwn, Chelabi:2015gpc}, as well as in 2+1 dimensions as discussed here. 

We have also shown in Fig. \ref{fig_dilaton} the dilaton profile used in this work with different values for the parameters $k$, $k_1$ and $k_2$. One sees that increasing $k$ will deepens the minimum of the dilaton profile. This implies that the overall scales also increase. In particular this increases the values of $B_c$ and $\sigma$. This suggests that increasing $k_1$ and $k_2$ will diminish the overall scales implying a decrease of $B_c$ and $\sigma$. 

\section*{}

\noindent {\bf Acknowledgments:} 
 D.M.R is supported by Conselho Nacional de Desenvolvimento Cient\'\i fico e Tecnol\'ogico (CNPq) and Coordena\c c\~ao de Aperfeiçoamento de Pessoal de N\' \i vel Superior (Capes) (Brazilian Agencies).  D.L. is supported by the National Natural Science Foundation of China (No.11805084) and the PhD Start-up Fund of Natural Science Foundation of Guangdong Province (No.2018030310457). H.B.-F. is partially supported by CNPq and Capes (Brazilian Agencies).


 \end{document}